\documentclass{article}
\usepackage[utf8]{inputenc}
\usepackage{amsmath,amssymb}
\usepackage{graphicx} 
\usepackage{xcolor}
\usepackage[backend=biber,style=numeric-comp,sorting=none]{biblatex}
\usepackage{appendix}
\usepackage{printlen}
\usepackage{authblk}
\usepackage{placeins}

\title{Wetting ridge dissipation at large deformations}

\author[1]{Martin H. Essink}
\author[2,3]{Stefan Karpitschka}
\author[4]{Hamza K. Khattak}
\author[4,5]{Kari Dalnoki-Veress}
\author[6]{Harald van Brummelen}
\author[1]{Jacco H. Snoeijer}
\affil[1]{Physics of Fluids Group, Mesa+ Institute, University of Twente, 7500 AE Enschede, The Netherlands.}
\affil[2]{Max Planck Institute for Dynamics and Self-Organization, 37077 G{\"o}ttingen, Germany.}
\affil[3]{Fachbereich Physik, Universit{\"a}t Konstanz, 78457 Konstanz, Germany.}
\affil[4]{Department of Physics and Astronomy, McMaster University, 1280 Main Street West, Hamilton, Ontario L8S 4M1, Canada.}
\affil[5]{UMR CNRS Gulliver 7083, ESPCI Paris, PSL Research University, 75005 Paris, France.}
\affil[6]{Department of Mechanical Engineering, Eindhoven University of Technology, 5600 MB Eindhoven, The Netherlands.}

\date{\today}

\begin{document}

\maketitle

\textit{
Liquid drops slide more slowly over soft, deformable substrates than over rigid solids. This phenomenon can be attributed to the viscoelastic dissipation induced by the moving wetting ridge, which inhibits a rapid motion, and is called ``viscoelastic braking". Experiments on soft dynamical wetting have thus far been modelled using linear theory, assuming small deformations, which captures the essential scaling laws. Quantitatively, however, some important disparities have suggested the importance of large deformations induced by the sliding drops. Here we compute the dissipation occurring below a contact line moving at constant velocity over a viscoelastic substrate, for the first time explicitly accounting for large deformations. It is found that linear theory becomes inaccurate especially for thin layers, and we discuss our findings in the light of recent experiments.
}

%

%%%%%%%%%%%%%%%%%%%%%%%%%%%%%%%%%%%%%%%%%%%%%%%%%%%%%%%%%%%%

%
\section{Introduction}
\begin{figure}[tb]
    \centering
	\includegraphics{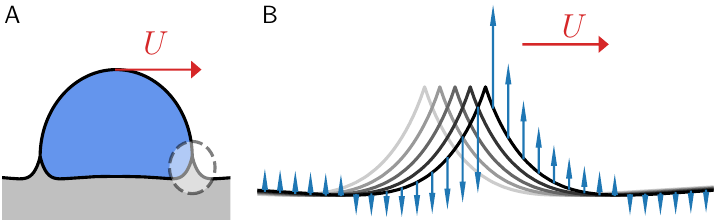}
	\caption{
 \textbf{(a)} Schematic view of a droplet moving to the right at velocity $U$, with the moving contact line indicated by the circle. \textbf{(b)} Zoom of the moving wetting ridge. The horizontal translation of the contact line induced a vertical motion of material points in the substrate, indicated by the blue arrows. This motion induces viscoelastic dissipation inside the substrate. (Note that the motion is not restricted to the interface but also occurs inside the bulk of the substrate, and is not strictly in the vertical direction.)
	}
	\label{fig:sketch}
\end{figure}
\begin{figure}[tb]
    \centering
	\includegraphics{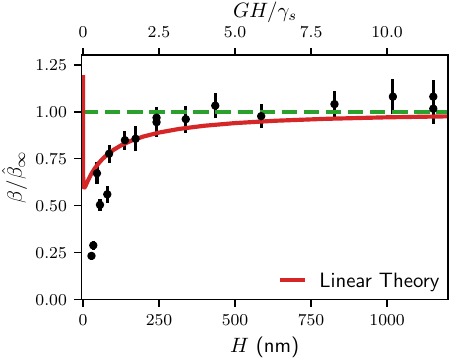}
	\caption{Dissipation $\beta$ in the wetting ridge, as a function of the substrate thickness $H$ and normalised thickness $GH/\gamma_s$. Experimental data is shown in black, and the corresponding linear viscoelastic theory in red. Recreated from data provided in \cite{khattak2022direct}.
	}
	\label{fig:beta_lec_exp}
\end{figure}

The motion of spreading and sliding drops offers a paradigmatic example of wetting dynamics. While relevant for any application where drops touch a wall, the wetting motion comes with challenges related to the fluid flow at small scales near the contact line \cite{GennesBrochard-WyartQuere2004,BEIM2009rmp,Snoeijer:ARFM2013}. It is of particular interest that the spreading and sliding is affected by deformability of the substrate~\cite{AnSn2020arfm,bico2018elastocapillarity,StDuPRL2013,SCPW2013potnaos,PWLL2014nc,PBDJ2017sm,Chakrabarti:L2018,Chen:COCIS2018,STSR2018nc,Xu:PRL2020,Coux:PNAS2020,Guan:PRL2020,Bardall:IJoAM2020,Zhao:NL2021,Cai:CM2021,Zhao:SM2022,Cai:AAPM2022,Nekoonam:AAMI2023,Hauer:PRL2023}. This field of research was pioneered by Shanahan and co-workers \cite{Shanahan:L1994,Carre:L1995,shanahan1995viscoelastic, CaGS1996n}, who discovered that contact lines move more slowly over rubbers and elastomers, as compared to rigid surfaces, regardless of the precise polymeric nature of the substrate. This slowing down was attributed to the viscoelastic dissipation inside the substrate that ensues when a wetting ridge translates over the substrate. 
The mechanism behind this so-called ``viscoelastic braking" is sketched in Figure~\ref{fig:sketch}. The surface tension of the droplet exerts a localized traction on the solid at the conctact line, creating a deformation of the underlying elastic substrate. Subsequently, a horizontal translation of this wetting ridge (red arrow) induces time-dependent displacements of material points (blue arrows). These displacements generate a strong viscoelastic dissipation inside the substrate that is responsible for the slowing down of the wetting motion \cite{Shanahan:L1994,Carre:L1995,shanahan1995viscoelastic, CaGS1996n}. Scaling theories were presented to compute the dissipation from the substrate's rheological properties \cite{LoAL1996lb,ZDNL2018pnasusa,AnSn2020arfm}, as a function of the speed of the contact line. In experiments, however, the dissipation, or rather the dissipative force, is usually not measured directly, but it is mostly quantified via a a change in the contact angle of the liquid-vapor interface with respect to the equilibrium angle. This change in angle was shown to originate from a rotation of the tip of the wetting ridge that, again, can be related to the viscoelastic response of the substrate \cite{KDGP2015nc,GKAS2020sm,Tamim:PRE2021,Tamim:JoFM2023,Oleron:a2023}. The  interpretations based on ``ridge rotation" and ``dissipation" offer strictly equivalent descriptions of dynamical wetting over soft substrates. This equivalence has been shown explicitly in the context of linear viscoelasticity, assuming small deformations, for which analytical predictions have indeed been obtained from the two different methods \cite{karpitschka2018soft_comment,GKAS2020sm}.
Recent experiments have for the first time provided direct measurements on the dissipation induced by moving contact lines \cite{khattak2022direct}. These measurements were achieved by dragging droplets over thin PDMS substrates by a flexible micro-pipette, which serves as a force sensor. The experiments confirmed the essential scaling laws of dissipation with contact line speed. Specifically, the dissipation per unit length of contact line, denoted $\mathcal D$, is of the form
\begin{equation}\label{eq:definitionbeta}
    \frac{\mathcal D}{\gamma U} =\beta(H)  \left(\frac{U}{U^*}\right)^n.
\end{equation}
In this expression $\gamma$ is the liquid-vapor surface tension, $U$ is the contact line velocity, $U^*$ is a characteristic viscoelastic velocity, and $n$ the rheological exponent of the substrate. The key element of interest here is the dependency on the substrate thickness, reported as a dimensionless factor $\beta(H)$. The experimental data for this factor $\beta$ obtained by  \cite{khattak2022direct}, is reproduced in Figure~\ref{fig:beta_lec_exp}. The data clearly show that dissipation decreases for smaller substrate thickness, and suggest that $\beta$ even vanishes for $H \to 0$. This result, however, is in stark disagreement with predictions from linear viscoelastic theory \cite{KDGP2015nc,GKAS2020sm,AlMo2021ijnme,Bardall:IJoAM2020}, superimposed as the red curve in Figure~\ref{fig:beta_lec_exp}. It was argued in \cite{khattak2022direct} that the mismatch between experiments and theory at small substrate thickness is due to the emergence of large deformations for wetting ridges on thin substrates. Namely, linear theory predicts the height of the wetting ridge $h_r$ to scale as $H^{3/4}$ in the limit of small substrate thickness \cite{ZDNL2018pnasusa}. This scaling for the ridge height implies a typical vertical strain $h_r/H \sim H^{-1/4}$, which obviously constitutes a very large deformation when $H$ becomes  small. As such, linear viscoelasticity does not offer a consistent theory of soft wetting on very thin substrates.
The goal of this paper is to model the statics and dynamics of wetting ridges, while consistently accounting for large viscoelastic deformations. First, we compute static wetting ridge profile using finite element simulations of Neo-Hookean solids, following a similar (macroscopic) scheme as in \cite{henkel2022soft}. The static results are presented in Section \ref{sec:static}, where we will focus on the ridge profiles in the limit of small substrate thickness. Then, we compute the dissipation following the idea sketched in Figure~\ref{fig:sketch}. The static profile is translated horizontally at constant velocity; subsequently, the dissipation is computed from the time-dependent deformations from this horizontal motion. In the context of small deformation theory, this procedure based on translation of a \emph{static} profile is known to give the exact \emph{dynamical} dissipation, to leading order in velocity \cite{KDGP2015nc}, and we show that this idea can be equally applied to large deformations. Namely, the change in the shape of the wetting ridge due to the motion offers only a higher order (in velocity) correction to the shape of the wetting ridge, and thus to the  dissipation. In Section~\ref{sec:translation} we formalise this procedure for large deformations, and the resulting dissipation factors $\beta(H)$ are presented in Section~\ref{sec:dissipationresults}. We also analytically derive the dissipation in the vicinity of the tip, which exhibits a power-law divergence. Importantly, we find that the dynamics-induced singularity is integrable, which justifies {\em a posteriori} the use of static profiles to compute the leading order dissipation.
In Section~\ref{sec:gent} we explore substrate constitutive relations beyond the Neo-Hookean model, specifically focusing on the effect of finite extensibility. We compare our results to the experiments in \cite{khattak2022direct} as discussed in the concluding Section~\ref{sec:conclusion}.

%

%%%%%%%%%%%%%%%%%%%%%%%%%%%%%%%%%%%%%%%%%%%%%%%%%%%%%%%%%%%%

%
\section{Static profiles}\label{sec:static}
\begin{figure}[tb]
    \centering
	\includegraphics{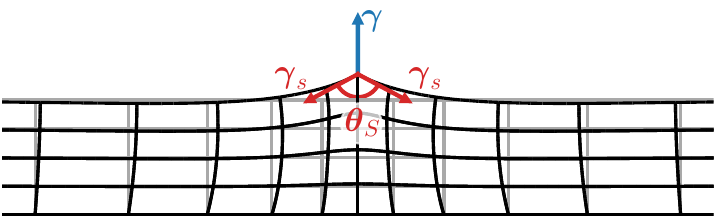}
	\caption{ Schematic representation of the wetting ridge on a solid substrate. The black lines indicate a course, non refined mesh in the deformed state. The mesh in reference state is shown in gray. The surface tensions $\gamma$ and $\gamma_s$ are shown in blue and red respectively. The solid opening angle $\theta_S$ is indicated between the surface tension arrows.
	}
	\label{fig:schematic}
\end{figure}

\subsection{Formulation}
Figure \ref{fig:schematic} shows a static numerical result. We consider a two-dimensional wetting ridge on a  substrate of finite thickness $H$. The substrate has a shear modulus $G$, and a surface energy $\gamma_s$ on the free surface. For simplicity we here consider $\gamma_s$ as constant, i.e.  independent of the stretch and the same on both sides of the wetting ridge. For asymmetric wetting conditions we refer to \cite{henkel2022soft}. The wetting ridge is induced by a point force $\gamma$ that pulls vertically on the surface, which models the force exerted by the liquid-vapor  interface at the contact line. 
The solid is deformed according to a mapping $\mathbf{x}=f(\mathbf{X})$, which maps the reference configuration $\mathbf{X}$ to the deformed configuration $\mathbf{x}$. The gradient of this mapping, the deformation gradient tensor $\mathbf{F} = \partial\mathbf{x}/\partial\mathbf{X}$, is a measure of the deformations of the solid. Restricting our considerations to solids with hyperelastic constitutive relations, from this tensor $\mathbf{F}$ an energy density $W(\mathbf{F})$ can be calculated, that represents the energy per unit area (in plane-strain conditions) that is stored in the elastic deformations of the solid. For most of this paper, we will consider the substrate to consist of an incompressible Neo-Hookean solid under plane-strain, for which the energy density reads
\begin{equation} \label{eq:W_Neo_Hookean}
    W(\mathbf{F}) = \frac{G}{2}\left( \mathrm{tr}\left(\mathbf{F} \cdot \mathbf{F}^T\right) - 2 \right) - p\left( \mathrm{det}(\mathbf F)-1 \right),
\end{equation}
where $G$ is the shear modulus of the material, and incompressibility is enforced through the Lagrange multiplier $p$. Integration of this energy density over the entire (undeformed) substrate yields the total mechanical energy stored in the substrate. With the addition of the surface energy $\gamma_s$ on the free surface, and the vertical point force $\gamma$ at $(X,Y)=(0,H)$, the total energy is
\begin{eqnarray}\label{eq:ftot}
    {\cal U} =%&=& 
    \int\limits dX \int\limits_{0}^{H} dY \,  W(\mathbf{F})
    + \int\limits \gamma_s \lambda_s dX
    + \gamma\, \mathbf{e}_y \cdot \mathbf{x}(0,H)
\end{eqnarray}
where $\lambda_s$ is the stretch of the free surface, and $\mathbf{e}_y$ the unit vector in vertical direction. The problem can be solved by minimisation of the energy functional $\cal U$ with respect to the deformation $\mathbf{x}$. 
In the final part of this paper we investigate the effect of finite extensibility of the polymeric substrate, which amounts to a modification of the elastic energy density $W(\mathbf F)$.
A typical result of the minimisation is shown in Figure \ref{fig:schematic}. The pulling force generates deformations throughout the entire substrate, which gives rise to both vertical and horizontal displacements. Immediately below the contact line, the wetting ridge is governed by an opening angle $\theta_S$ dictated by the Neumann balance of the three surface tensions \cite{PAKZ2020prx}. Therefore, the solid opening angle $\theta_S$ and the ratio of surface energies $\gamma/\gamma_s$ are directly related as
\begin{equation} \label{eq:theta_s}
    \theta_S = 2\arccos\left(\frac{1}{2}\frac{\gamma}{\gamma_s}\right) 
    \quad \Longleftrightarrow \quad
    \frac{\gamma}{\gamma_s} = 2 \cos \left(\frac{\theta_S}{2}\right).
\end{equation}
Hence, we shall be using $\theta_S$ as the measure that quantifies the relative strength of the surface tensions. We also remark that the theory of linear elasticity is only expected to be valid for $\theta_S$ close to $180^\circ$, i.e. $\gamma/\gamma_s \ll 1$. Experimentally, however, this condition is typically not fulfilled, which supports our interest of performing simulations with large deformation theory. The ratio of surface tension to elasticity is expressed via the elastocapillary length; here it can be defined as both $\gamma/G$ and $\gamma_s/G$. Roughly speaking, the former sets a typical vertical scale,  while the latter is a horizontal scale of deformation. As a dimensionless measure of the (inverse) substrate thickness, we will be using $\gamma_s/GH$. The opening angle $\theta_S$, or equivalently the ratio $\gamma/\gamma_s$, provides the second dimensionless parameter in the problem.
%

%%%%%%%%%%%%%%%%%%%%%%%%%%%%%%%%%%%%%%%%%%%%%%%%%%%%%%%%%%%%

%
\subsection{Numerical method}
The minimisation of the total energy \eqref{eq:ftot} is implemented using the open-source finite element framework Nutils \cite{ZZVF2020nutils}. 
We start with an unrefined mesh of $4\times 192$ elements, representing only the right half of the wetting ridge, for symmetry reasons. 
A high aspect ratio is achieved by increasingly stretching the elements when moving further away form the contact line, as can be observed in Figure \ref{fig:schematic}. The mapping $X\mapsto X(1+X)$ produces a mesh of width $2352 H$, which is wide enough to be considered infinite for the purposes of this study. The elements are hierarchically refined up to $18$ times using a residual-based adaptive method, such that an element size of $\Delta < 10^{-4}\ \gamma_s/G$ is achieved near the contact line. The bottom of the mesh is constrained in both horizontal and vertical direction, imitating a substrate bonded to a rigid surface. The right boundary is also constrained in both directions. The left boundary represents the symmetry axis of the wetting ridge, thus the deformations are only constrained in horizontal directions, with no shear in the vertical direction. The top surface representing the free surface is traction-free, with only stresses resulting from surface tension.
For the simulation of sharp wetting ridges a continuation approach was required. In this case the surface energy ratio $\gamma/\gamma_s$ was increased in a number of smaller steps, such that a value up to $\gamma/\gamma_s=1.5$ ($\theta_S = 82.8^\circ$) could be reached, before discretisation artifacts typical for sharp kinks at free surfaces inhibit convergence of the solver. 

%%%%%%%%%%%%%%%%%%%%%%%%%%%%%%%%%%%%%%%%%%%%%%%%%%%%%%%%%%%%

%
\begin{figure}[tb]
    \centering
	\includegraphics{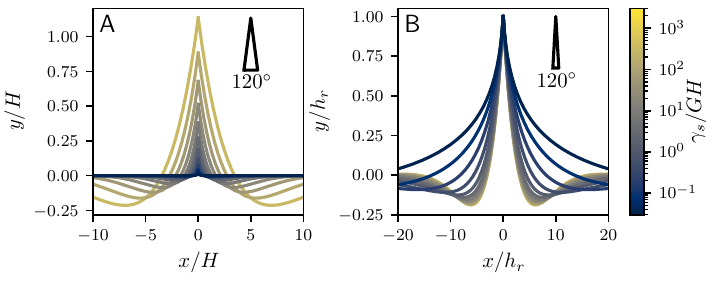}
	\caption{Numerical free surface profiles for various values of the elastocapillary length, with lines colored according to the colorbar on the right. A surface energy ratio $\gamma/\gamma_s=1.0$ was used, corresponding to a solid angle $\theta_S = 120^\circ$. \textbf{(a)} Normalised by the thickness of the undeformed solid layer $H$. \textbf{(b)} Normalised by the height of the wetting ridge $h_r$, which leads to a collapse of the data in the soft limit.
	}
	\label{fig:profiles}
\end{figure}
\subsection{Effect of substrate thickness}
We are interested in studying how the wetting ridge is affected when varying the substrate thickness. Specifically, we wish to investigate the limit of thin substrates, for which linear elasticity becomes inconsistent since large strains are predicted \cite{ZDNL2018pnasusa,khattak2022direct}. In the simulations we thus vary $\gamma_s/GH$, which represents the ratio of the elastocapillary length $\gamma_s/G$ and the substrate thickness $H$. Figure \ref{fig:profiles}a shows free surface profiles of the wetting ridge for a range of $\gamma_s/GH$, with the horizontal and vertical axis normalized by $H$. The presented data are for $\gamma/\gamma_s=1$, so that $\theta_S=120^\circ$. The blue data corresponds to the limit of small elastocapillary length, or equivalently of large substrate thickness. In this limit, the deformation (normalized by $H$) appears as a small feature on top of the comparatively thick substrate. The yellow data correspond to the opposite limit of large elastocapillary length, or small thickness. Now, the elastic deformation becomes large compared to the layer thickness, and continues to grow upon increasing $\gamma_s/GH$. Since the opening angle remains constant between the curves, the profiles all appear to exhibit a similar shape in the soft limit, only changing in scale. We examine whether the shape of the wetting ridge is indeed truly self-similar in the thin limit, by scaling the profiles by the height of the wetting ridge $h_r$, both vertically and horizontally. The resulting set of curves is displayed in Figure \ref{fig:profiles}b. Indeed, the wetting ridge profiles converge to a universal shape for thin substrates ($\gamma_s/GH \gg 1$), suggesting that the ridge height -- rather than the substrate thickness -- is the relevant length scale in the thin limit.
\begin{figure}[tb]
    \centering
	\includegraphics{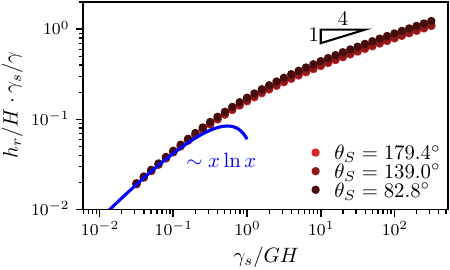}
	\caption{Height of the wetting ridge as a function of the elastocapillary number. The height of the wetting ridge is scaled by the ratio of surface energies $\gamma/\gamma_s$ to overlay results for multiple solid opening angles. The blue line and the triangle show, respectively, the $\sim x\ln x$ regime for thick layers and the $1/4$ power law for thin layers, predicted by linear theory. % for thin layers\cite{ZDNL2018pnasusa}.
	}
	\label{fig:hr}
\end{figure}
The remaining task is to understand the scaling of the ridge height $h_r$ as a function of $\gamma_s/GH$. The numerical results for for the ridge height are reported in Figure \ref{fig:hr}, for different values of $\theta_S$. In the plot the ridge height $h_r/H$ is normalized by an additional factor $\gamma_s/\gamma$, which leads to a nearly perfect collapse of the data for all $\theta_S$. The Figure clearly reveals the presence of two regimes. For thick substrates we find a scaling that is close to linear, and consistent with the prediction from two-dimensional linear elasticity $h_r \sim \gamma/G \ln (GH/\gamma_s)$ \cite{Lima2012epje,LWBD2014jfm}.

Our main interest, however, lies in the limit of thin substrates, which pertains to the data on the right in Figure \ref{fig:hr}. The ridge height $h_r$ now closely follows a scaling exponent $1/4$. This behavior is perfectly in line with predictions based on linear elasticity, which suggest the wetting ridge to grow as \cite{ZDNL2018pnasusa} 
\begin{equation} \label{eq:lin_ridgeheight}
    \frac{h_r}{H} \sim \frac{\gamma}{\gamma_s} 
    \left(\frac{\gamma_s}{GH}\right)^{1/4}.
\end{equation}
Hence, the wetting ridge continues to grow, without any evidence for a bound, upon increasing the elastocapillary length. Importantly, since $h_r/H$ serves as a measure for strain, at least at the location of the ridge, we clearly see that strains do not remain small: they even diverge in the limit $H\to 0$. However, the results from the Neo-Hookean solid (see Figure \ref{fig:hr}) are perfectly in line with the prediction from linear theory \eqref{eq:lin_ridgeheight}. Hence, even though linear elasticity is no longer consistent in the limit of thin substrates, it correctly predicts the horizontal and vertical scaling of the wetting ridge on Neo-Hookean solids.
%

%%%%%%%%%%%%%%%%%%%%%%%%%%%%%%%%%%%%%%%%%%%%%%%%%%%%%%%%%%%%

%
\section{From static profiles to dissipation}\label{sec:translation}
We now turn to the main focus of the paper, which is to compute the dissipation inside a wetting ridge as a function of its translation velocity. For this we employ the procedure outlined in the Introduction. In the spirit of the sketch in  Figure~\ref{fig:sketch}, we consider a horizontal translation of the full static solution of Figure~\ref{fig:schematic} at a constant velocity, and compute the dissipation from the resulting motion of material points. In this section we present the technical details on how to perform this calculation while consistently accounting for large deformations. The results of this calculation, in particular the spatial density of dissipation and the behavior of the global dissipation factor $\beta(H)$ as defined in (\ref{eq:definitionbeta}), will be presented in section~\ref{sec:dissipationresults}.
%

%%%%%%%%%%%%%%%%%%%%%%%%%%%%%%%%%%%%%%%%%%%%%%%%%%%%%%%%%%%%

%
\subsection{Stress relaxation at large deformations}
The theory of linear viscoelasticity can be expressed using a stress relaxation function that captures the memory of stress under time-dependent deformations. For small deformations, this formalism is typically written in scalar form as
\begin{equation}\label{eq:linearVE}
    \sigma(t) = \int_{-\infty}^t dt' \, \psi(t-t') \dot \gamma(t').
\end{equation}
This equation gives a relation between (a component of) the stress and the history of the rate-of-strain  $\dot \gamma$. The function $\psi(t-t')$ is called the stress relaxation function, and defines the linear rheological response of the material. Since we wish to deal with dissipation in the presence of large deformations, we need to use the appropriate tensorial form of this relation. 
In the general description of nonlinear viscoelastic solids, the stress at time $t$ is a functional over the entire history of deformation \cite{wineman2009nonlinear}. The deformation must thus be described using a time-dependent mapping $\mathbf x = f(\mathbf X,t)$. We assume that the system is in the (stress-free) reference configuration at time $t_0$, such that $f(\mathbf X,t_0)=\mathbf X$. The history of deformation can then be described by the two-time Finger tensor $\mathbf B(t,t')$, also known as the left Cauchy-Green deformation tensor, which compares the deformation of the material between two different times, namely the current time $t$ and a time in the past $t' < t$. Formally, this quantity is defined from the time-dependent deformation tensor
\begin{align}
    \mathbf{F}(t,t') = \frac{\partial \mathbf{x}(t)}{\partial \mathbf{x}(t')} \quad \Rightarrow \quad 
    \mathbf{B}(t,t') &= \mathbf{F}(t,t') \cdot \mathbf{F}^{T}(t,t').
\end{align}
The Finger tensor with respect to the reference configuration, as it appears in static elasticity, is recovered as $\mathbf B(t,t_0)$.

In what follows we will restrict ourselves to isotropic models of solids with ``finite linear viscoelasticity"; this terminology expresses that the models are not restricted to small deformations, but are still taken as linear functional of $\mathbf B$. Specifically, we shall be using the form \cite{wineman2009nonlinear}
\begin{equation}\label{eq:fullVEmodel}
    \boldsymbol{\sigma}(t) = \boldsymbol{\sigma}^{\rm el}(t) + 
    \int_{-\infty}^t dt' \, \left[ - \psi_1(t-t') \frac{\partial \mathbf B(t,t')}{\partial t'} 
    + \psi_2(t-t') \frac{\partial \mathbf B(t,t')^{-1}}{\partial t'} \right].
\end{equation}
The first term $\boldsymbol{\sigma}^{\rm el} $(t) is the elastic stress that, as usual, depends on $\mathbf B(t,t_0)$ and $\mathbf B(t,t_0)^{-1}$ that measures the deformation with respect to the reference configuration at $t=t_0$. The dissipative, viscoelastic part of the stress is expressed by the integral, where the kernels $\psi_{1,2}$ decay to zero in the limit of large $t-t'$; this condition ensures that the elastic response is recovered after the configuration is held constant (i.e. after stopping the ``flow"). In general, the relaxation functions $\psi_{1,2}$ can be chosen to depend on $\mathbf B(t,t_0)$, which would render the functional nonlinear and the model beyond ``finite linear viscoelasticity"; for simplicity this possibility will not be considered here. Since the Neo-Hookean solid, and also the Gent model that will be discussed later, do not explicitly depend on $\mathbf B(t,t_0)^{-1}$, we will make the further simplification that $\psi_2=0$. A final reduction can be achieved for the Neo-Hookean solid, which is defined by the elastic stress $\boldsymbol{\sigma}^{\rm el} (t)= \mathbf G (\mathbf B(t,t_0)-\mathbf I)$. Setting the reference time $t_0=-\infty$, the elastic stress can be absorbed inside the integral as:
\begin{equation}\label{eq:hetislaat}
    \boldsymbol{\sigma}(t) =  -\int_{-\infty}^t dt' \,\ \psi(t-t') \frac{\partial \mathbf B(t,t')}{\partial t'}.
\end{equation}
Here we made use of the property that $\mathbf B(t,t)=\mathbf I$, and we defined the stress relaxation function
\begin{equation}
\psi(t-t')= G + \psi_1(t-t').
\end{equation}
We remind that $\psi_1 \to 0$ in the long-time limit, such that $\psi \to G$ ensuring a solid-like behavior.

Upon comparing to (\ref{eq:linearVE}), we find that (\ref{eq:hetislaat}) forms a tensorial viscoelastic rheology that admits a consistent description for large deformations. The explicit manipulation of this form, however, is intricate as it involves tensor products of the mapping $\mathbf F(t,t')$. A great simplification is obtained when expressing the tensors using a curvilinear formulation. In this formulation, material points are described by curvilinear coordinates $q^i$. The infinitesimal distance $ds$ between two nearby points $q^i$ and $q^i+dq^i$ follows, using Einstein summation convention, as 
\begin{equation}
ds^2 = g_{ij} dq^i dq^j,
\end{equation}
where $g_{ij}$ is the so-called metric tensor. For a time-dependent deformation, the distance between material coordinates changes over time, such that the metric must be time-dependent $g_{ij}(t)$. In Appendix~\ref{apx:to_metric}, we demonstrate that covariant components of the the rate-of-strain tensor $\dot{\boldsymbol \gamma}$ are simply given by $\dot g_{ij}(t)$, so that $\dot{\boldsymbol \gamma}$ is naturally connected to the time-derivative of the metric. Likewise, one can show that the contravariant components at time $t$ of the Finger tensor $\mathbf B(t,t')$ are given by $g^{ij}(t')$, i.e. involving the metric at time $t'$. With these expressions, the rheology (\ref{eq:hetislaat}) takes on a remarkably simple form (Appendix \ref{apx:to_metric}),  
\begin{equation}\label{eq:sigmacurvilinear}
    \sigma^{ij}(q,t) = -  \int_{-\infty}^t dt' \, \psi(t-t') \dot g^{ij}(q,t').
\end{equation}
This form is identical to linear viscoelasticity, as given by (\ref{eq:linearVE}), but now offering a proper tensor form admitting large deformations; the geometric nonlinearities associated to large deformations are implicitly accounted for by the metric tensor. For transparency, we have introduced in (\ref{eq:sigmacurvilinear})  the spatial coordinates $q$ as arguments of both stress and metric. We recall that these are associated with material points, so that the memory integral is to be performed at a constant material point. 
%

%%%%%%%%%%%%%%%%%%%%%%%%%%%%%%%%%%%%%%%%%%%%%%%%%%%%%%%%%%%%

%
\subsection{Dissipation for a traveling wave}
\subsubsection{Density of mechanical work}
With an explicit expression for the viscoelastic stress at hand, we can evaluate the power density,
\begin{equation}
\mathcal P=\frac{1}{2}\boldsymbol{\sigma}: \dot{\boldsymbol{\gamma}},
\end{equation}
 giving the volumetric density of work done by the stress per unit time. Let us note that under the standing assumption of incompressibility, the power density is identically defined in the reference and current configurations. 
 The power density does not give direct access to the dissipation within the solid, as it is comprised of a reversible (elastic) contribution and an irreversible (dissipative) contribution. Hence, the work can be split as
\begin{equation} \label{eq:work_split}
    \mathcal P = \frac{D \mathcal W}{D t} + \epsilon.
\end{equation}
In this expression $\mathcal W$ is the reversibly stored elastic energy, while the remainder $\epsilon$ is the irreversibly dissipated power per unit volume. The term $D\mathcal W/Dt$ represents the material derivative of the stored energy, {\em i.e.} the time derivative with respect to a fixed position in the reference configuration.

To illustrate these concepts, let us consider a passing contact line, moving above a certain material point inside the viscoelastic substrate. The total work per unit volume by the passing wave is obtained by integrating the power over all times,
\begin{equation}
    \int\limits_{-\infty}^\infty \mathcal P\; dt = \mathcal W\bigg|_{t=-\infty}^{t=\infty} + \int\limits_{-\infty}^\infty \epsilon\; dt,
\end{equation}
Prior and after the passage of the contact line, there is no elastic energy stored. Hence, the term involving $\mathcal W$ drops out, so that the \emph{integral} over the power gives the total amount of dissipated energy. So even though $\mathcal P \neq \epsilon$, the total dissipation can be computed by considering an integral over the power density.

\subsubsection{Travelling wave solution}

We will now develop explicit expressions for the dissipation below a moving contact line. For this, we first write the power density in curvilinear form 
\begin{align}
\mathcal P &= \frac{1}{2}\boldsymbol{\sigma}(t) : \boldsymbol{\dot{\gamma}}(t) 
= \frac{1}{2}  \sigma^{ij}(X,Y,t) \dot g_{ij}(X,Y,t) \nonumber \\  
&= - \frac{1}{2} \int\limits_{-\infty}^t d\tau  \tilde \psi\left(\frac{t-\tau}{\lambda}\right) \dot g^{ij}(X,Y,\tau) \dot g_{ij}(X,Y,t) .
\end{align}
In the first line we made explicit that we will use the horizontal and vertical of reference positions as the curvilinear coordinates, i.e. $q^1=X$ and $q^2=Y$. In the second line we inserted the expression for $\sigma^{ij}$ in terms of the stress relaxation function $\psi$, as derived in the preceding section. In order to keep track of dimensions, we made the argument of the relaxation function dimensionless with the relaxation time scale $\lambda$, according to $\tilde \psi(t/\lambda)=\psi(t)$.
The power density can thus be written in terms of time-derivatives of the metric tensor, evaluated at times $t$ and $t'$. For a contact line moving to the right as a travelling wave, we find that all fields are of the form $g(X,Y,t) = \mathcal{G}(X-Ut,Y)$. Correspondingly, the power density due to a travelling wave can be expressed as 
\begin{equation}
\mathcal P(X,Y,t) = -\frac{U}{2} \int\limits_{-\infty}^t d(U\tau)  \tilde \psi\left(\frac{t-\tau}{\lambda}\right)  \mathcal{G}^{ij\prime}(X-U\tau,Y) \mathcal{G}_{ij}'(X-Ut,Y),
\end{equation}
where a prime indicates a derivative with respect to the first argument of $\mathcal G$. Making the transformation of variables $\hat X = X-Ut$ and $\tilde  X = X-U\tau$, one obtains
\begin{equation} \label{eq:P_psi}
\mathcal P(\hat X,Y) = -\frac{U}{2}  \int\limits_{\hat X}^{\infty} d\tilde X  \tilde \psi\left(\frac{\tilde X-\hat X}{U \lambda}\right)  \mathcal{G}_{ij}'(\hat X,Y) \mathcal{G}^{ij\prime}(\tilde X,Y).
\end{equation}
Hence, for a travelling wave solution, the history-dependence at a co-moving point $\hat X$ can be inferred by looking at the deformations at $\tilde X > \hat X$, i.e. points that have already passed $\hat X$. As explained in the discussion of Figure~\ref{fig:sketch}, to leading order in velocity we can evaluate this expression for the work by taking the fields $\mathcal{G}_{ij}(X,Y)$ and $\mathcal{G}^{ij}(X,Y)$ directly from the static numerical solutions obtained in Section~\ref{sec:static}. 

The central goal of our study is to compute the total  dissipation $\mathcal D$ in the substrate, representing the total energy dissipated per unit time per unit length of contact line. This quantity is obtained by integrating $\epsilon$ over the entire domain of the substrate. Given that for a travelling wave the total stored elastic energy does not change in time, it is only displaced, we simply find that 
\begin{align}\label{eq:dissipation_integral}
\mathcal D &= \int\limits_{0}^H dY \int\limits_{-\infty}^\infty dX\; \epsilon =  \int\limits_{0}^H dY \int\limits_{-\infty}^\infty dX\; \mathcal P.
\end{align}
Hence, the integral of the power density over the entire domain directly offers the total amount of dissipation.

\subsubsection{Density of dissipation} \label{sec:dissipation_density}

While we shall be using (\ref{eq:dissipation_integral}) to compute the \emph{global} dissipation inside the substrate, it is still of interest to consider the \emph{local} volumetric density of dissipation $\epsilon$. To compute this quantity, we follow \cite{snoeijer2020relationship} and define a conformation tensor $A^{ij}$. The conformation tensor characterizes the local degree of stretching, and can be viewed as the order parameter that describes the thermodynamic state of the system. Given the linear constitutive law (\ref{eq:sigmacurvilinear}), it is natural to define a linear relation between stress and the order parameter, according to $\sigma^{ij}=GA^{ij}$. This is a simple ``Neo-Hookean" relation between stress and the order parameter, for which the corresponding reversible work can be described by a linear stored energy density 

\begin{equation}
\label{eq:neoHookean}
\mathcal W=\frac{1}{2}G g_{ij}A^{ij}. 
\end{equation}
To appreciate that this is the correct form of the energy, we take its time-derivative,
\begin{equation}\label{eq:split}
\frac{D\mathcal W}{Dt} = \frac{1}{2} G A^{ij} \dot{g}_{ij} 
+  \frac{1}{2} G g_{ij} \dot{A}^{ij}.
\end{equation}
We recall that the material derivative in this expression reduces to regular time derivatives, since here the fields are expressed as functions of the curvilinear material coordinates. Noting that $\dot{g}_{ij}$ represent the covariant components of the tensor $\dot{\boldsymbol{\gamma}}$, we recognize the first term on the right as the power density by internal stress, $\frac{1}{2}\boldsymbol{\sigma}: \dot{\boldsymbol{\gamma}}$. From this we can make the identification that $\sigma^{ij}= G A^{ij}$, consistent with our initial definition of the conformation tensor. Given that $D\mathcal W/Dt$ represents the reversible part of the work, the second term on the right thus represents the irreversible dissipation $\epsilon$, i.e. 
\begin{equation} \label{eq:epsilon_def}
\epsilon = - \frac{1}{2} G g_{ij} \dot{A}^{ij} = -\frac{1}{2} g_{ij} \dot{\sigma}^{ij}.% = \mathrm{Tr} \stackrel{\triangledown}{\mathbf \sigma}.
\end{equation}
For a thermodynamically consistent constitutive relation, the expression for $\epsilon \geq 0$, which we will indeed find in our numerical results. 
%

%%%%%%%%%%%%%%%%%%%%%%%%%%%%%%%%%%%%%%%%%%%%%%%%%%%%%%%%%%%%

%
\subsection{Chasset-Thirion rheology}
Until now we have not considered any specific relaxation function $\psi$. In line with previous work, we will consider the Chasset-Thirion rheology \cite{thirion1967viscoelastic,khattak2022direct}, that has a complex modulus
\begin{equation} \label{eq:CT_fourier}
    G'+iG'' = G\left[1+(i\omega \lambda)^n \right],
\end{equation}
with $0< n < 1$, and where $\lambda$ offers the characteristic relaxation timescale. This complex modulus indeed provides an excellent fit of typical PDMS substrates used in soft wetting experiments \cite{KDGP2015nc,GKAS2020sm}. 
On the time-domain, this model is recovered to a stress relaxation function
\begin{equation}
    \tilde \psi(\bar t) = G \left[ 1 + \Gamma(1-n)^{-1} \bar t^{-n} \right],
\end{equation}
where $\bar t = t/\lambda$ and the result further involves the $\Gamma$-function. In the expression for the power density, the constant term of the relaxation function expresses the elastic stress, and will not give a resultant contribution to the total dissipation integral. To compute $\mathcal D$, it is thus sufficient to retain only the time-dependent part of the relaxation function, so that the power density can be written as
\begin{equation} \label{eq:dissipation_fin_dim}
\frac{\mathcal P}{UG} = -\frac{1}{2} \frac{(U\lambda )^n }{ \Gamma(1-n)} \int\limits_{X}^{\infty} d\tilde X \, \, 
\frac{\mathcal{G}_{ij}'(X,Y) \mathcal{G}^{ij\prime}(\tilde X,Y)}{\left(\tilde X-X\right)^n}  .
\end{equation}
This expression can also be used to compute the power density of a purely elastic material, i.e. $\psi=G$, upon setting $n=0$.

%

%%%%%%%%%%%%%%%%%%%%%%%%%%%%%%%%%%%%%%%%%%%%%%%%%%%%%%%%%%%%

%
\subsection{Non-dimensionalisation}
As a final step, we wish to express the total dissipation in the form of equation (\ref{eq:definitionbeta}):
the velocity-dependence is pulled out explicitly and $\beta$ is a dimensionless factor that reflects the dependence of the dissipation on the substrate thickness \cite{khattak2022direct}. In the presented dimensionless formulation of the problem, $\beta$ depends on the parameters $\gamma_s/(GH)$ and $\gamma/\gamma_s$. Since the total dissipation follows from an integral of power density over the entire substrate, we analogously define 
\begin{equation} \label{eq:alpha_def}
    \frac{\mathcal P}{U\gamma/H^2}  = \left(U/U^* \right)^n  \alpha.
\end{equation}
where $\alpha$ reflects the thickness-dependent part of the power density. The integral relation between $\mathcal D$ and $\mathcal P$ as given by (\ref{eq:dissipation_integral}) then conveys  

\begin{equation}
\beta = \int_0^1 d\bar Y \int_{-\infty}^\infty d\bar X \, \alpha,
\end{equation}
where $\bar X=X/H$ and $\bar Y=Y/H$.  
 
 With the definition of the Chasset-Thirion rheology in place, the exponent $n$ introduced in (\ref{eq:definitionbeta}) is now well-defined. In addition, the characteristic horizontal velocity $U^*=\gamma_s/(G\lambda)$ naturally emerges from the elastocapillary length $\gamma_s/G$ and the relaxation time $\lambda$. Substituting the above scalings in (\ref{eq:dissipation_fin_dim}), we find the dimensionless power density 
\begin{equation} \label{eq:localwork_nondim}
\alpha = -\frac{1}{2}  \frac{\gamma_s}{\gamma} \left(\frac{\gamma_s}{GH}\right)^{n-1} \frac{1}{ \Gamma(1-n)} 
\int\limits_{\bar X}^{\infty} d\bar{\tilde X} \, \, 
\frac{\mathcal{G}_{ij}'(\bar X,\bar Y) \mathcal{G}^{ij\prime}(\bar{\tilde X},\bar{Y})}{\left(\bar{\tilde X}-\bar X\right)^n}  .
\end{equation}
Integration over the entire substrate finally gives,
\begin{equation} \label{eq:beta_num}
\beta = -\frac{1}{2}\frac{\gamma_s}{\gamma} \left(\frac{\gamma_s}{GH}\right)^{n-1}\frac{1}{ \Gamma(1-n)}\int\limits_{0}^1 d\bar Y   \int\limits_{-\infty}^\infty d \bar X \int\limits_{\bar X}^{\infty} d\bar{\tilde X} \, \, 
\frac{\mathcal{G}_{ij}'(\bar X,\bar Y) \mathcal{G}^{ij\prime}(\bar{\tilde X},\bar Y)}{\left(\bar{\tilde X}-\bar X\right)^n}.
\end{equation}
This expression gives the sought after dissipation factor $\beta$, which is a function of $\gamma_s/GH$ and $\theta_S$ (via the factor $\gamma/\gamma_s$). We recall that also the static deformation profiles are encoded in $\mathcal{G}_{ij}(\bar X,\bar Y)$ and $\mathcal{G}^{ij}(\bar X,\bar Y)$, as obtained from the finite element simulations, depend on the parameters $\gamma_s/GH$ and $\theta_S$.
\subsubsection{Benchmark: The limit of small deformations}
The large-deformation result for $\beta$ will be evaluated numerically, and compared to the analytical expression for the limit of small deformations. The latter consists of a single integral, rather than of a triple integral, and reads \cite{GKAS2020sm,khattak2022direct}:
\begin{eqnarray} \label{eq:beta_lin}
    \hat \beta &=& -\frac{\gamma}{GH}\left( \frac{\gamma_s}{GH}\right)^n \sin \left(\frac{n \pi}{2} \right)
    \int_{-\infty}^\infty \frac{dQ}{2\pi} \frac{K(Q) Q^{n+1}}{\left(1+\frac{\gamma_s}{GH} Q^2 K(Q)\right)^2},
\end{eqnarray}
where $K(Q)$ is the elastic kernel for a dimensionless layer of finite thickness,
\begin{equation}
    K(Q) = \frac{1}{2Q}\frac{\sinh{(2Q)}-2Q}{\cosh{(2Q)}+2Q^2+1}.
\end{equation}
In this expression it is explicit that the dissipation integral $\beta$ is only a function of $\gamma_s/(GH)$ and $\gamma/\gamma_s$. In the limit of infinite thickness ($\gamma_s/GH\ll1$), the integral can be performed analytically and one finds the reference value for a half-space
\begin{equation}
    \hat{\beta}_\infty = \frac{\gamma}{\gamma_s} \frac{n\, 2^{n-1}}{\cos \left(\tfrac{n\pi}{2}\right)}.
\end{equation}
Analogously to $\hat{\beta}_\infty$, we also define $\beta_\infty$ for the large-deformation case as the limiting value of $\beta$ for infinite thickness ($\gamma_s/GH\ll1$).
\section{Dissipation results}\label{sec:dissipationresults}
\begin{figure}[tb]
    \centering
	\includegraphics{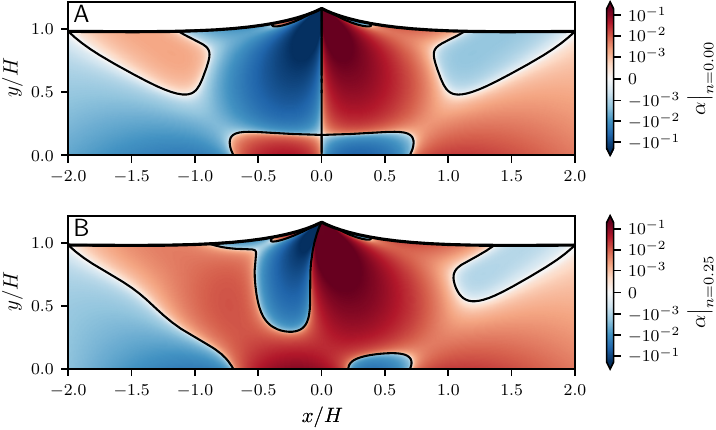}
	\caption{Dimensionless power density $\alpha$ in the substrate below a wetting ridge moving to the right.  Positive contributions (red) indicate mechanical energy being stored into the solid, where the negative contributions (blue) indicate that energy is released. The black contours emphasise the transitions between positive and negative power density. Results are obtained for an opening angle of $\theta_S=120^\circ$ and an elastocapillary number of $\gamma_s/GH = 1.0$. \textbf{(a)} A purely elastic solid response ($n=0$)  exhibits an anti-symmetric distribution of work in the solid. \textbf{(b)} The viscoelastic contribution to the power for a typical for the case $n=0.25$. In this case, the symmetry is broken and the power density will integrate to nonzero global dissipation.
	}
	\label{fig:dissipation}
\end{figure}
\begin{figure}[tb]
    \centering
	\includegraphics{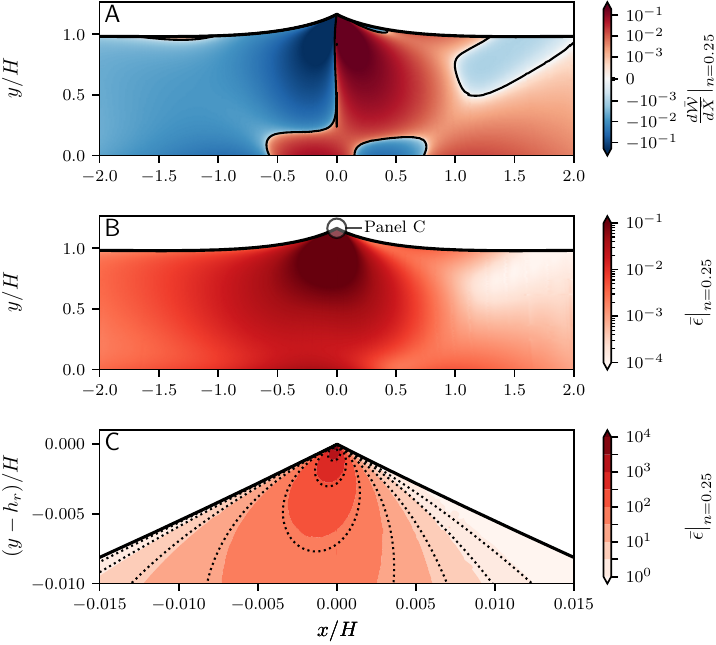}
	\caption{The viscoelastic power as given in Figure \ref{fig:dissipation}b is split into \textbf{(a)} Reversible power $dW/dt$ and \textbf{(b)} Irreversible power $\epsilon$, the latter being the density of dissipation. The plotted quantities are made dimensionless using the same scaling as $\alpha$. \textbf{(c)} Zoom of the dissipation density near the tip, comparing numerics (colors) to the asymptotic analysis (dashed lines).
	}
	\label{fig:dissipation2}
\end{figure}
In the previous section we have shown how to compute dissipation within the viscoelastic substrate. We now present the results of this calculation, while systematically varying  $\gamma_s/GH$ and $\theta_S$.
\subsection{Spatial distribution of dissipation inside the solid}

\subsubsection{Numerics}
Before turning to the total dissipation induced by a moving ridge, we first investigate where this dissipation occurs inside the substrate. More precisely, we compute the dimensionless power density $\alpha$, as defined by \eqref{eq:alpha_def}, where the result is computed from the explicit formula \eqref{eq:localwork_nondim}. We recall that even though this quantity integrates up to the total dissipation inside the wetting ridge, it is not a measure of ``local dissipation". This becomes immediately clear when $\alpha$ is plotted for a purely elastic solid, for which no dissipation can occur. This elastic case is obtained from  \eqref{eq:localwork_nondim} 
 with $n=0$, and the resulting power density is given in Figure \ref{fig:dissipation}a. Red zones are regions of positive work, where energy is stored in the solid; primarily upon arrival of the ridge. Blue zones are regions of negative work, where energy is released; primarily after the passage of the ridge. For the purely elastic case of Figure \ref{fig:dissipation}a, the distribution of work is perfectly anti-symmetric, such that all the stored energy is eventually released. The resulting power density therefore integrates up to zero; as should be the case for for a purely elastic substrate that exhibits no viscoelasticity.
On the contrary, this left-right symmetry is broken for viscoelastic solids. This can be seen in Figure \ref{fig:dissipation}b, where we show the viscoelastic contribution to the power density, as computed from  \eqref{eq:localwork_nondim} with $n\neq 0$. Specifically, we consider $n=0.25$, which is close to the experimental value $n=0.23$ in \cite{khattak2022direct}. The storage of energy (red) ahead of the wetting ridge still closely resembles that of the purely elastic substrate. However, much less of this energy is released (blue) in the wake of the ridge. Upon integrating over the entire substrate, one thus obtains a resultant global (positive) dissipation.
It is of interest to disentangle the viscoelastic power of Figure \ref{fig:dissipation}b, into the reversible part ($d\mathcal W/dt$) and the irreversible part ($\epsilon$), using the splitting according to (\ref{eq:work_split}). The results are shown in Figure \ref{fig:dissipation2}; the plotted quantities are made dimensionless using the same scalings as $\alpha$. The upper panel (a) corresponds to the reversible part of the viscoelastic power, $d\mathcal W/dt$, which is again computed by applying the travelling wave method to the reversible energy defined by (\ref{eq:neoHookean}).  Interestingly, this reversible  viscoelastic power density is not left-right symmetric, even though (by definition) its integral  adds up to zero. The actual dissipation density $\epsilon$ is reported in Figure \ref{fig:dissipation2}b. This result is obtained by subtracting the reversible power from the total power, as explained in (\ref{eq:work_split}), and is strictly non-negative. From Figure \ref{fig:dissipation2}b we observe that the dissipation is largest close to the contact line, where the rate of deformation is largest. Also, we find that the maximum dissipation rate is skewed to the left, which is the part of the substrate where the contact line has already passed. A detailed zoom of the local dissipation is provided in Figure \ref{fig:dissipation2}c, where it is compared to an analytical solution derived below using an asymptotic analysis.

\subsubsection{Asymptotics: Zooming in near the tip}

Close to the contact line, at scales well below the elastocapillary length, a static wetting ridge can be described as a fold of and elastic half-space to an angle $\theta_S$ \cite{PAKZ2020prx}. Using polar coordinates ($R,\Phi$) to describe the reference configuration, the half-space corresponds to the domain $-\pi \leq \Phi \leq 0$ with $R \geq 0$. This is mapped onto a deformed configuration, with polar coordinates $(r,\phi)$, according to \cite{singh1965note}
\begin{equation}
    (r,\phi) = \left(\frac{R}{\sqrt{b}}, 
    b\left(\Phi + \frac{\pi}{2}\right) - \frac{\pi}{2}\right).
\end{equation}
This transformation amounts to a uniform compresssion by a factor $b=\theta_S/\pi$ in the azimuthal direction, which is centerred around $\Phi=-\pi/2$ (and $\phi=-\pi/2$). This radial stretching then automatically follows to ensure incompressibility. Indeed, as reported in \cite{PAKZ2020prx}, numerical solution of the wetting ridge indeed closely follows this fold map in the asymptotic limit where $r\to 0$.

\begin{figure}[tb]
    \centering
    \includegraphics{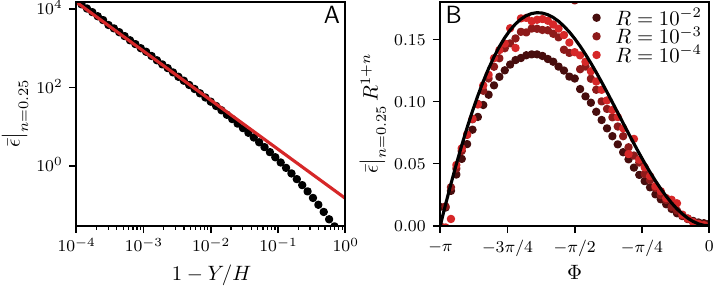}
    % \textcolor{red}{\uselengthunit{in}\printlength{\textwidth}}
	\caption{Asymptotic solution for the dissipation near the tip (solid lines), compared to interpolated numerical data (symbols) without any adjustable parameters. \textbf{(a)} The density of dissipation (taken close to the angle of maximum dissipation, at $\Phi=-2.0$) as a function of the distance to the tip. Simulations confirm the power-law divergence with exponent $1+n$. \textbf{(b)} Normalized angular dependence of the dissipation density. Numerical results converge to the analytical prediction as the distance to the tip decreases.
	}
	\label{fig:asymptotics}
\end{figure}

To compute the dissipation, we can proceed by following the formalism as described above. Rather than using the numerical static profiles, we now use the analytical solution to compute the dissipation asymptotically close to the tip. Details of this calculation are given in Appendix~\ref{apx:asymptotics}, where we obtain a closed form analytical expression. For convenience, the result is presented in dimensional units, and reads: 
\begin{equation}\label{eq:epsilonasymptotic}
    \epsilon(R,\Phi) =  \frac{1}{2} G \lambda^n 
    \frac{\left(b^2-1\right)^2}{b^2} 
    \frac{\pi n(n+1)}{\Gamma(1-n) \sin (n \pi)} 
    \sin(\Phi) \sin (n\Phi)
    \left(\frac{U}{R}\right)^{n+1}.
\end{equation}
Figure \ref{fig:dissipation2}c provides a direct comparison between this result and the numerical evaluation of the dissipation. 
Both the dependencies on $R$ and in $\Phi$ provide an excellent description of the numerical dissipation profiles, as is evident from the direct comparison in Figure~\ref{fig:asymptotics}.

We thus find that the dissipation density diverges as $1/R^{n+1} \sim 1/r^{n+1}$, upon approaching the tip. The limit of linear viscoelasticity is recovered by taking $1-b \ll 1$, corresponding to angles close to $\theta_S=\pi$. It is clear that linear viscoelasticity, as used previously in \cite{karpitschka2018soft_comment,GKAS2020sm}, correctly captures the scaling exponents for the dissipation. In fact, the scaling can be understood by considering the loss modulus, $G'' \sim G(\omega \lambda)^n$, in conjunction with a typical frequency, $\omega \sim U/r$. The dissipation density then follows as 
\begin{equation} \label{eq:arfm_epsilon}
    \epsilon 
    \sim G'' \cdot \omega 
    \sim G (\omega \lambda)^n \omega
    \sim G  \lambda^n (U/r)^{n+1} ,
\end{equation}
in perfect agreement with the exact result (\ref{eq:epsilonasymptotic}). An important consequence of this scaling is that, since $0<n<1$, this dissipation singularity is integrable at small scale -- this can be seen by multiplying $\epsilon$ with the area element $r d\theta dr \sim d\theta dr/r^n $, which gives a singularity weaker than $1/r$; see also the discussion in \cite{GKAS2020sm}. This ensures that the total dissipation, as reported below, indeed remains finite. This also justifies {\em a posteriori} our approach to compute the dissipation based on static profiles. Note that the inner region of soft wetting, close to the contact line, is described by a wedge geometry~\cite{PAKZ2020prx,singh1965note}, which is less singular than the Flamant problem~\cite{malkov2008analysis,WLJH2018sm}. The description proposed here avoids the intricacies of expansions~\cite{DeRL2020sm}, and clearly shows that large deformations do not affect the nature of the singularity: the scaling laws for the dissipation density are the same as those found in linear theory. We remark that these asymptotic results cannot be used to predict the global dissipation, since the large-scale features of the ridge act as cut-off for the inner asymptotics.

\subsection{Global dissipation}
\begin{figure}[bt]
    \centering
	\includegraphics{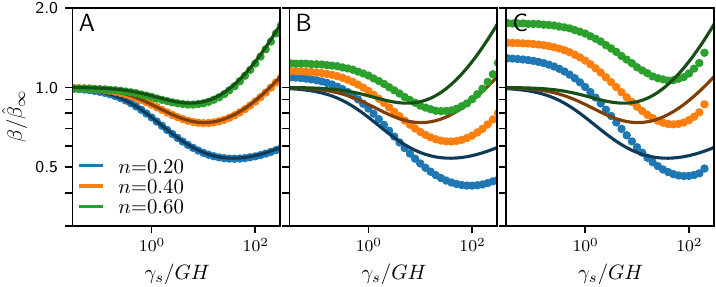}
	\caption{Normalised dissipation in the wetting ridge as a function of the elastocapillary length for various values of the parameter $n$ in the Chasset-Thirion rheology. Results from linear theory are shown as solid lines, where the numerical results are denoted by the symbols. The ratio of surface energies for the numerical results are
	\textbf{(a)} $\theta_S=179.4^\circ$, \textbf{(b)} $\theta_S=120.0^\circ$, \textbf{(c)} and $\theta_S=82.8^\circ$.
	}
	\label{fig:beta_lec_n}
\end{figure}

We now turn to the global dissipation, which is expressed using the dissipation factor $\beta$ that is defined in the introduction as equation (\ref{eq:definitionbeta}). We compute this factor in the presence of large deformations, using \eqref{eq:beta_num}, and the results will be systematically compared to the small-deformation linear theory \eqref{eq:beta_lin}. This direct comparison is shown in the three panels of Figure \ref{fig:beta_lec_n}, each corresponding to a different opening angle $\theta_S$ of the wetting ridge. Naturally, one would expect the dissipation to match the linear theory in the case of $\theta_S$ close to $180^\circ$, as the deformations in this limit are small. Indeed, for an angle $\theta_S=179.4^\circ$ in Figure \ref{fig:beta_lec_n}a the simulations almost perfectly match the small deformation theory, for the full range of $\gamma_s/GH$. This excellent quantitative agreement (including the normalisation factor for thick layers $\hat{\beta}_\infty$, which was not fitted but taken directly from linear theory), offers a validation of our numerical scheme.

\begin{figure}[tb]
    \centering
	\includegraphics{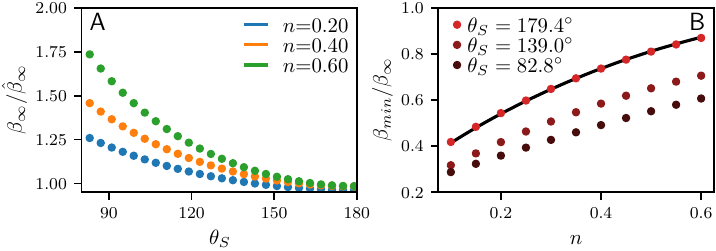}
	\caption{ \textbf{(a)} Dissipation coefficient for thick substrates $\beta_\infty$ as a function of the opening angle $\theta_S$ for various values of the parameter $n$, normalised by the value $\hat{\beta}_\infty$ predicted from linear elasticity. \textbf{(b)} Minimum value of the dissipation coefficient $\beta$ as a function of the parameter $n$, for a number of surface energy ratios $\gamma/\gamma_s$.
	}
	\label{fig:beta_min}
\end{figure}
As the wetting ridge becomes sharper, with smaller $\theta_S$, some deviations become apparent. In Figures \ref{fig:beta_lec_n}b and \ref{fig:beta_lec_n}c we see that with a decreasing opening angle (or increasing $\gamma/\gamma_s$) two important changes occur. 
First of all, in the limit of infinitely thick layers ($\gamma_s/GH \to 0$), the dissipation coefficients $\beta \to \beta_\infty$ deviates from the linear prediction $\hat \beta_\infty$. 
This deviation depends on the parameter $n$, as can be seen in Figure \ref{fig:beta_min}a. Secondly, for thinner substrates another change in the $\beta$ curves is observed. It seems the minimum value of $\beta$, here called $\beta_{min}$ decreases when the wetting ridge becomes sharper. The minimum $\beta_{min}$ is further quantified in Figure \ref{fig:beta_min}b. It is clear that the minimum of the dissipation decreases significantly when the solid angle $\theta_S$ takes on smaller values. 
\begin{figure}[bt]
    \centering
	\includegraphics{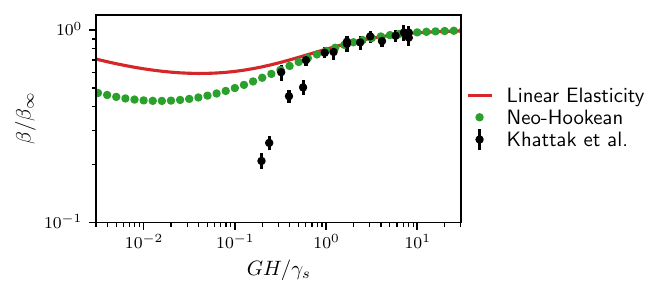}
	\caption{Dissipation in the wetting ridge as a function of the substrate thickness, normalised by the elastocapillary length. Dissipation is normalised by $\beta_\infty$, the value of $\beta$ observed for thick substrates. The experimental results from \cite{khattak2022direct} are overlaid in black. A surface energy ratio $\gamma/\gamma_s=1.0$ was used for the numerical results, corresponding to an internal solid angle $\theta_S = 120^\circ$. The value $n=0.25$ is chosen to approximately match the experimental data.
	}
	\label{fig:beta_lec}
\end{figure}
Typical experimental values for $\theta_S$ lie around $90^\circ$. Our results thus imply that significant quantitative differences between experiment and linear theory are to be expected. On the other hand, the qualitative trends of linear theory remain intact: in all cases the dissipation exhibits a minimum, and does not tend to zero in the limit of small thickness. This is in contrast with the experimental data, suggesting that $\beta$ vanishes in the limit $H \to 0$. To further elaborate on this, we replot the experimental results of \cite{khattak2022direct} in Figure \ref{fig:beta_lec}, compared to both linear and large deformation results. Note that $\beta$ is now normalised by $\beta_\infty$, since the large thickness value of $\beta$ is experimentally used as a fitting parameter, and the horizontal axis displays the dimensionless thickness $HG/\gamma_s$ (rather than its inverse in the preceding plots). The large deformation results are clearly a closer match to the experiments; still, these do not capture the strong decrease in $\beta$ that was observed for experiments on thin layers.
%

%%%%%%%%%%%%%%%%%%%%%%%%%%%%%%%%%%%%%%%%%%%%%%%%%%%%%%%%%%%%

%
\section{Finite extensibility}\label{sec:gent}
\subsection{Modeling considerations}
Until now we have only considered the Neo-Hookean constitutive behavior, augmented with a Chasset-Thirion viscoelastic relaxation. While quantitative differences appeared with respect to the linear theory, we found that qualitative trends for the ridge height and the dissipation are the same, even in the limit of very thin substrates. We now explore whether qualitative changes appear for constitutive relations with a more strongly nonlinear behavior, beyond the Neo-Hookean solid.
Soft solids such as the PDMS used in the experiments \cite{khattak2022direct}, typically consist of a cross-linked network of polymer chains. While the polymers have some freedom to deform, the polymer chains cannot be stretched indefinitely (in contrast to a Neo-Hookean solid). A model that captures the finite extensibility of the polymers is the Gent solid model, which imposes limiting behavior on the strain invariant $I=\mathrm{tr}\left(\mathbf{F} \cdot \mathbf{F}^T\right) - 2$, such that it cannot exceed a limiting factor $I_m$. Since $I$ can be seen as a ``mean stretch", $I_m$ translates to a direct limit on the stretch, with a maximum uniaxial stretch $\lambda_m$ defined by $I_m=\lambda_m^2 + \lambda_m^{-2} - 2$. The Gent solid model has a free energy ~density\cite{gent1996new}
\begin{equation} \label{eq:W_Gent}
    W(\mathbf{F}) = -\frac{G\ I_m}{2} \ln \left( 1 - \frac{
     \mathrm{tr}\left(\mathbf{F} \cdot \mathbf{F}^T\right) - 2 }{I_m}
    \right)
    - p\left( \mathrm{det}(\mathbf F)-1 \right),
\end{equation} 
which develops a diverging stress when $I \to I_m$. In the limit of $I/I_m \to 0$, however, the model converges to the Neo-Hookean solid \eqref{eq:W_Neo_Hookean}. Therefore, for moderate deformations the two models should be nearly indistinguishable. For large deformations, as is the case for thin substrates, qualitative differences are anticipated.
Once again, the dissipation can be computed from static solutions, by using the travelling wave Ansatz. For the Gent model, the starting point of the viscoelastic analysis is no longer \eqref{eq:hetislaat} but for the more general relation  \eqref{eq:fullVEmodel}, with a modified expression for $\boldsymbol \sigma^{\rm el}$. However, the elastic part of the stress does not contribute to the dissipation. So if we use for $\psi_1$ the time-dependent part of the Chasset-Thirion relaxation function, we are still allowed to use the very same expression for the dissipation factor $\beta$, namely equation  (\ref{eq:beta_num}). The effect of finite extensibility is then encoded via the change of the static profiles.
\subsection{Results}
\begin{figure}[tb]
    \centering
	\includegraphics{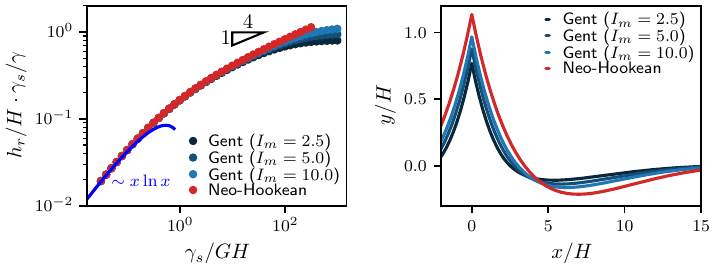}
	\caption{\textbf{(a)} Height of the wetting ridge as a function of the elastocapillary number, similar to Figure \ref{fig:hr} but including finite extensibility results for a number of $I_m$ values. The blue line and the triangle show, respectively, the $\sim x\ln x$ regime for thick layers and the $1/4$ power law for thin layers, predicted by linear theory \cite{ZDNL2018pnasusa}. A surface energy ratio $\gamma/\gamma_s=1.0$ was used for all simulations, corresponding to an internal solid angle $\theta_S = 120^\circ$. \textbf{(b)} The corresponding surface profiles for a relatively thin substrate ($\gamma_s/GH=3.16\times10^2$). 
	}
	\label{fig:hr_gent}
\end{figure}
Since the finite extensibility only comes into play at relatively large deformations, it is expected that differences between the Neo-Hookean model and Gent models will only be apparent at large deformations. This is indeed the case when plotting the height of the wetting ridge as a function of the elastocapillary number, as done in Figure \ref{fig:hr_gent}a. Only for very thin substrates, where the wetting ridge becomes relatively large, we see a qualitative difference between the two models (see also Figure \ref{fig:hr_gent}b). Specifically, the Gent model predicts a saturation to the unbounded growth of the ridge height; leading to a breakdown of the 1/4 scaling law. This implies that pulling more strongly on the solid does not lead to larger deformations when reaching the finite extensibility limit. The smaller the value of $I_m$, the smaller the maximum height of the ridge.
\begin{figure}[tb]
    \centering
	\includegraphics{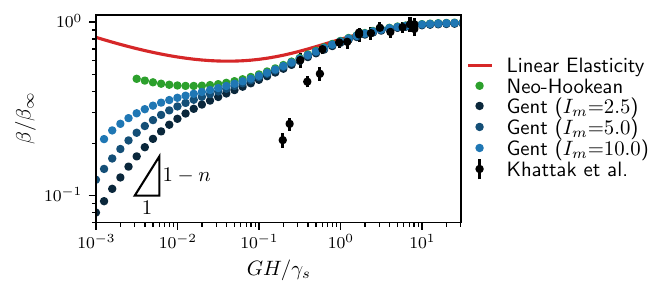}
	\caption{Dissipation in the wetting ridge as a function of the substrate thickness, similar to Figure \ref{fig:beta_lec}. The data for the finite extensibility model is added in blue. All simulations use a surface energy ratio $\gamma/\gamma_s=1.0$, or an internal solid angle $\theta_S = 120^\circ$, and $n=0.25$ in the rheology. The expected power law \eqref{eq:gent_beta_scaling} is displayed in the bottom left.
	}
	\label{fig:beta_gent}
\end{figure}
\sloppy The saturation of the deformation also impacts the total dissipation. Roughly speaking, the finite extensibility means that the metric factors appearing in the dissipation integral \eqref{eq:beta_num} are also subjected to a saturation, even if one continues to increases $\gamma_s/GH$. 
By virtue of the finite extensibility of the metric tensors in \eqref{eq:beta_num}, as well as their derivatives, are bounded independent of $H$. Expression \eqref{eq:beta_num} then yields the following estimate for $\beta$: 
\begin{equation} \label{eq:gent_beta_scaling}
    \beta\lesssim \left(\frac{\gamma_s}{GH}\right)^{n-1},
\end{equation}
whenever finite extensibility comes into play. Given that $n<1$, this implies $\beta \lesssim H^{1-n} \to 0$ in the limit of vanishing substrate thickness. Hence, the Gent model could offer a qualitative change in the behavior of $\beta$, with a trend that is in line with experimental observations.
This hypothesis is confirmed in Figure \ref{fig:beta_gent}, where we report $\beta$ for the Gent model. In contrast to the Neo-Hookean model and the linear theory, the Gent model does not exhibit a minimum, but gives a vanishing dissipation as $HG/\gamma_s \to 0$. The observed scaling is in very good agreement with the prediction  (\ref{eq:gent_beta_scaling}). This reveals that strong nonlinearities are indeed important to explain the experimental observations. Having said that, the Gent model in combination with a ``linear" Chasset-Thirion model is not able to provide a quantitative match of the experiment on the thinnest substrates. The precise value of $I_m$ in the experiments depends sensitively on the elastomer type and crosslinking density. In the experimental results a value $I_m \sim 10$ would be a reasonable estimate \cite{milner2017creasing}, which is within the captured numerical range.  
%

%%%%%%%%%%%%%%%%%%%%%%%%%%%%%%%%%%%%%%%%%%%%%%%%%%%%%%%%%%%%

%
\section{Conclusion}\label{sec:conclusion}
In this study we set out to find an explanation for the apparent disagreement between the dissipation of moving wetting ridges in experiments and linear theory that was observed in \cite{khattak2022direct}.  It was found that linear theory does not capture the monotonic decrease of dissipation when reducing the thickness of the substrate;   instead, linear theory predicts a minimum in dissipation followed by a sharp increase  for very thin layers. It was hypothesised that the appearance of a minimum was an artefact of the linear solid model, and would disappear for nonlinear solid models. To test this we performed finite element simulations of the wetting ridge, allowing for the consideration of nonlinear solid models. From these simulation results the dissipation is then calculated.
We find that the free surface profiles converge to a self-similar shape in the limit of thin substrates. This means that even though the wetting ridge grows relative to the substrate thickness, its shape remains unaltered, but is only scaled up. Remarkably, the wetting ridge in this limit, normalised by the layer thickness, shows the same power law growth behavior as predicted by linear elasticity \cite{ZDNL2018pnasusa,Oleron:a2023}. Unexpectedly, even in the limit of large deformation the growth of the wetting ridge persists.
Knowing the shape of the wetting ridges and deformation within the elastic layer allows us to calculate the rate of work in the substrate. By integration, we can then find the dissipation of a moving wetting ridge. Our results show that the global dissipation for opening angles near $180^\circ$ is indeed in perfect agreement with the predictions of linear theory; the convergence validates our numerical method. For sharper wetting ridges we start to see effects of nonlinear elasticity, as would be expected for the larger deformations associated with the sharper ridges. The dissipation for thick substrates increases, but the minimum and asymptote for thin layers  persist. 
Since the molecular structure of typical soft materials is not fully captured by a Neo-Hookean solid, we also considered a finite extensibility model. Such a model incorporates the limited extensibility of the polymer chains that make up the bulk. We find that on thick substrates, finite extensibility does not significantly impact the size of the wetting ridges, while on thin substrates it sets a maximum height. Dissipation calculations show an asymptote that goes to zero in the thin substrate limit. So, the introduction of strong nonlinearities beyond the Neo-Hookean solid are clearly of importance to capture the essentical feature of dynamical wetting experiments on very thin layers. However, the Gent model combined with linear viscoelastic dissipation does not yet offer a quantitative match to the experimental data. Future studies could be dedicated to add the nonlinear stress of the Gent model into the dissipation model, to see if a quantitative match can be achieved. Conversely, experiments on thinner layers combined with macroscopic nonlinear rheological characterisations (beyond $G'/G''$) could be of interest for future work. 

\FloatBarrier
\appendix
\section{Deformation in terms of metric tensor} \label{apx:to_metric}
It will turn out convenient to express the relevant deformation tensors using curvilinear material coordinates $\{ q^i\}$, with $i=1,2$ for our two-dimensional problem. We therefore provide the connection between the curvilinear formulation and the quantities that can be extracted from the numerical results. The discussion below is adapted from  \cite{green1992theoretical,snoeijer2020relationship}.
The numerical simulations make use of the deformation gradient tensor $\mathbf{F}=\partial \mathbf x/\partial \mathbf X$. We now consider a dynamical problem, for which the mapping $\mathbf x=f(\mathbf X,t)$ becomes time-dependent. For this reason we from now on use the notation $\mathbf x(t)$ and replace $\mathbf X = \mathbf x(t_0)$, where, without loss of generality, we consider the system to be in its reference configuration at $t=t_0$. With this, $\mathbf F$ and its inverse $\mathbf{F}^{-1}$ can be expressed as
\begin{align}
    \mathbf{F}(t) &= \frac{\partial \mathbf{x}(t)}{\partial \mathbf{x}(t_0)}
    = \frac{\partial \mathbf{x}(t)}{\partial q^i} \frac{\partial q^i}{\partial \mathbf{x}(t_0)}
    = \mathbf e_i(t) \otimes \mathbf E^i,\\
    \mathbf{F}^{-1}(t) %= \mathbf{F}(t_0,t') 
    &= \frac{\partial \mathbf{x}(t)}{\partial \mathbf{x}(t_0)} = \frac{\partial \mathbf{x}(t_0)}{\partial q^i} \frac{\partial q^i}{\partial \mathbf{x}(t)} = \mathbf E_i \otimes \mathbf e^i (t),
\end{align}
where we introduced the time-dependent curvilinear basis vectors 
\begin{equation}
    \mathbf e_i(t) = \frac{\partial \mathbf x(t)}{\partial q^i}, \quad  \mathbf e^i(t) = \frac{\partial q^i}{\partial \mathbf x(t)},
\end{equation}
and the time-independent reference basis vectors 
\begin{equation}
    \mathbf E_i = \mathbf e_i(t_0), \quad  \mathbf E^i = \mathbf e^i(t_0).
\end{equation}
In what follows we need to compute the history of deformation by comparing states at two different times, say the current time $t$ and a time in the past $t' < t$. The corresponding mapping $\mathbf F(t,t')$ is defined by
\begin{align}
    \mathbf{F}(t) \cdot \mathbf{F}^{-1}(t') \nonumber
    &= \mathbf e_i(t) \otimes \mathbf E^i \cdot \mathbf E_j \otimes \mathbf e^j(t') \\
    &= \mathbf e_i(t) \otimes \mathbf e^i(t')
     = \frac{\partial \mathbf{x}(t)}{\partial \mathbf{x}(t')} = \mathbf{F}(t,t').
\end{align}
When evaluating the associated Finger tensor, we obtain
\begin{align}
    \mathbf{B}(t,t') &= \mathbf{F}(t,t') \cdot \mathbf{F}^{T}(t,t')  \nonumber \\
    &= \mathbf{F}(t) \cdot \mathbf{F}^{-1}(t') \cdot \mathbf{F}^{-T}(t') \cdot \mathbf{F}^T(t) \nonumber \\
    &= \mathbf e_i(t) \otimes \mathbf e^i(t') \cdot \mathbf e^j(t') \otimes \mathbf e_j(t) \nonumber \\
    &= g^{ij}(t') \, \mathbf e_i(t) \otimes \mathbf e_j(t).\label{eq:BTT}
\end{align}
Here we introduced the contravariant metric tensor $g^{ij}$, which is defined as
\begin{equation}
g^{ij}(t) = \mathbf e^i(t) \cdot \mathbf e^j(t), \quad g_{ij}(t) = \mathbf e_i(t) \cdot \mathbf e_j(t),
\end{equation}
and for completeness we also gave the covariant metric $g_{ij}$. It follows from (\ref{eq:BTT}) that the two-time Finger tensor is obtained by expressing the contravariant metric at time $t'$ on the curvilinear basis at time $t$. Similarly the inverse of this tensor can be evaluated 
\begin{align}
    \mathbf{B}^{-1}(t,t') &= \mathbf{F}^{-T}(t) \cdot \mathbf{F}^{T}(t') \cdot \mathbf{F}^{}(t') \cdot \mathbf{F}^{-1}(t) \nonumber \\
    &= g_{ij}(t') \mathbf e^i(t) \otimes \mathbf e^j(t). \label{eq:Bi}
\end{align}
Naturally, $g_{ij}(t')$ and $g^{ij}(t')$ act as one-anothers inverse, being related respectively to $\mathbf B^{-1}$ and $\mathbf B$.
The constitutive relations of viscoelastic materials involve time-derivatives of the two-time Finger tensor. These can be evaluated as 
\begin{align}
    \frac{\partial\mathbf{B}(t,t')}{\partial t'} 
    &= \dot{g}^{ij}(t') \, \mathbf e_i(t) \otimes \mathbf e_j(t) \nonumber\\
    &= \mathbf{F}(t) \cdot \left(
    \dot{\mathbf{F}}^{-1}(t') \cdot \mathbf{F}^{-T}(t') + \mathbf{F}^{-1}(t') \cdot \dot{\mathbf{F}}^{-T}(t')
    \right) \cdot \mathbf{F}^T(t), \label{eq:dB_dt}\\
    \frac{\partial\mathbf{B}^{-1}(t,t')}{\partial t'} 
    &= \dot{g}_{ij}(t') \, \mathbf e^i(t) \otimes \mathbf e^j(t) \nonumber\\
    &= \mathbf{F}^{-T}(t) \cdot \left(
    \dot{\mathbf{F}}^{T}(t') \cdot \mathbf{F}^{}(t') + \mathbf{F}^{T}(t') \cdot \dot{\mathbf{F}}^{}(t')
    \right) \cdot \mathbf{F}^{-1}(t), \label{eq:dBi_dt}
\end{align}
providing the expression in both in symbolic notation and in curvilinear components. 
Now we see the great advantage of the curvilinear formulation. When expressed using the current basis vectors $\mathbf e_i$ (resp. $\mathbf e^i$), the time-derivatives of $\mathbf B$ (resp. $\mathbf B^{-1}$) reduce to ordinary time-derivatives of the metric components in the past $\dot g^{ij}$ (resp. $\dot g_{ij}$). By consequence, the constitutive relations take on comparatively simple forms when expressed in curvilinear coordinates. Phrased differently, the curvilinear description can be seen as an elegant way to effectuate the intricate projections between current and past configurations.

For completeness, we also provide the following connections (see \cite{snoeijer2020relationship} for details):
\begin{align}
    \dot{\mathbf F} = (\nabla \mathbf v)^T \cdot \mathbf F, &&
    \dot{\mathbf F}^{-1} = - \mathbf F^{-1} \cdot (\nabla \mathbf v)^T .
\end{align}
We thus obtain,
\begin{align}
    \frac{\partial \mathbf{B}^{-1}(t,t')}{\partial t'} 
    &= \dot{g}_{ij}(t') \, \mathbf e^i(t) \otimes \mathbf e^j(t)\nonumber \\
    &= \mathbf{F}^{-T}(t) \cdot 
    \mathbf F(t')^T \cdot \left(\nabla \mathbf v(t') + (\nabla \mathbf v)^T(t') \right)
     \cdot \mathbf{F}^{}(t') \cdot \mathbf{F}^{-1}(t)\nonumber\\
    &= \mathbf{F}^{-T}(t,t')  \cdot \dot{\boldsymbol \gamma}(t') \cdot   \mathbf{F}^{-1}(t,t')
\end{align}
The time-derivative of $\mathbf B^{-1}$ thus reflects the shear-rate $\dot {\boldsymbol \gamma}=\nabla \mathbf v + (\nabla \mathbf v)^T$ in the past, but projected on the current basis. When setting $t=t'$ after taking the derivative, we find
\begin{align}
    \left.\frac{\partial \mathbf{B}^{-1}(t,t')}{\partial t'} \right|_{t'=t}
    &= \dot{g}_{ij}(t) \, \mathbf e^i(t) \otimes \mathbf e^j(t)= \nabla \mathbf v + (\nabla \mathbf v)^T = 
     \dot{\boldsymbol{\gamma}}(t)
\end{align}
So, $\dot g_{ij}(t)$ combined with the basis at the current time directly gives  $\dot{\boldsymbol \gamma}(t)$.

\section{Asymptotics} \label{apx:asymptotics}
We now compute the dissipation near the contact line, assuming the deformation is described by a fold map. The reference domain is an elastic halfspace ($Y<0$), which is flat initially, which in polar coordinates reads
\begin{equation}
    (X,Y) \to (R,\Theta) = (\sqrt{X^2+Y^2}, \arctan(Y/X)),
\end{equation}
where $R$ and $\Theta$ are the radial and tangential coordinates respectively. The incompressible fold map in the absence of surface shear can now be defined as \cite{singh1965note}
\begin{equation}
    (r,\theta) = \left(\frac{R}{\sqrt{b}}, b\Theta+\frac{(b-1)\pi}{2}\right).
\end{equation}
where $r$ and $\theta$ are the counterparts of $R$ and $\Theta$ in the deformed configuration, $b=\theta_s/\pi$ and one confirms that incompressibility is indeed satisfied ($rdrd\theta = RdRd\Theta$). The deformed configuration in Cartesian coordinates can be expressed as,
\begin{equation}
    (r,\theta) \to (r \cos\theta, r \sin\theta) = (x,y),
\end{equation}
from which, the deformation gradient tensor $\mathbf{F}=d\mathbf x/d\mathbf X$ can be computed. Then, the metric quantities $g_{ij}(t')$, $\dot g^{ij}(t')$ and $\dot g_{ij}(t')$ can be calculated from the expressions \eqref{eq:Bi}, \eqref{eq:dB_dt} and \eqref{eq:dBi_dt} respectively. With the metric expressions in hand, it is now possible to directly compute dissipation density \eqref{eq:epsilon_def}. Using Mathematica, we obtain \eqref{eq:epsilonasymptotic} that is successfully compared to numerics.

\FloatBarrier

\section*{Acknowledgements}
We thank Bruno Andreotti for discussions. M.H.E. and J.H.S. acknowledge funding from NWO through VICI Grant No. 680-47-632, S.K. from the German research foundation (DFG, project no. 422877263). K.D.V. and H.K.K. acknowledge financial support from the Natural Science and Engineering Research Council of Canada, and HKK acknowledges funding from the Vanier Canada Graduate Scholarship.

\printbibliography

\end{document}